# Human Navigation Behaviour and Brain Dynamics in Real-world Contexts

Pablo Fernandez Velasco[1], Antoine Coutrot[2] and Hugo J. Spiers[3*]

1. Centre for the Sciences of Place and Memory, University of Stirling, UK
2. LIRIS, CNRS, INSA Lyon, Universite Claude Bernard Lyon 1, Lyon, France
3. Institute of Behavioural Neuroscience, Department of Experimental Psychology, Division of Psychology and Language Sciences, University College London, UK

For correspondence: * h.spiers@ucl.ac.uk

**Abstract**

The study of navigation behaviour and the associated brain dynamics have been a focus increasing research over the last decades. Coinciding with this has been an increased focus on a more ecological understanding of cognition. Here we review recent research seeking to provide a more naturalistic, ecological understanding of human navigation behaviour and brain dynamics. Research in this area falls into four categories: testing navigation in real-world environments, analysis of data collected from tracking individuals during daily life, navigation in simulated or virtual environments mimicking the real-world, and mobile brain recording methods. Combining these different approaches to understand the neural basis of navigation shows excellent promise. We conclude with future directions for this research area.

**Introduction**

There is a growing effort towards increasing ecological validity in the cognitive sciences. This effort stems from the limitations associated with studying cognition in the lab under tightly controlled conditions that might differ from real world settings in important ways. The progress towards more ecologically valid experiments has been facilitated by theoretical, methodological and analytical developments [1–6]. One of the domains in which ecological validity is particularly important is spatial navigation. How humans find their way spans a wide array of cognitive processes that are adapted to specific environments [7]. Navigation engages capacities such as memory, attention, path integration, planning and mental imagery [8–10], and the strategies and abilities involved vary with respect to demographic, geographic and cultural factors [11–13]. How all of these factors interact in naturally occurring settings can be lost in laboratory settings. Here, we review recent experimental efforts to study real world navigation. We divide this research into four main categories: 1) real-world behavioural testing, 2) virtual reality (VR) testing in environments that mimic real-world environments, 3) mobile tracking of participants collected outside of experiments in everyday life, and 4) mobile brain recording. We identify the strengths of each approach, the key insights, and their limitations.

## 1. Real-world testing of navigation

The most straightforward way to increase the ecological validity of experiments is to conduct experiments both in the lab and in a similar real-world context [14–24]. This is a way of ensuring that lab-based and/or online experiments are predictive of real-world performance and that effects that are found inside the lab also occur outside it. Real-world experiments are extremely useful in showing how people navigate in situations that they daily encounter. Testing real-world behaviour with neuropsychological patients can also give important insights into the brain regions that are involved in navigation [25]. Nevertheless, it is also useful to use simulated or virtual environments because these provide consistency across participants, can allow manipulation of environmental variables impossible in the real-world, and can be deployed to test larger numbers of participants, addressing questions that would be impossible or lack statistical power with small samples [12,26]. A key question with the development of testing environments is: does it predict navigation in the real-world?

In an effort to create a standardised battery of navigational tests that were ecologically valid, Bonavita and colleagues [22] compared performance in a laboratory-based and in an ecological environment for tasks including route, landmark and survey knowledge, and found a significant correlation in all tasks except for route knowledge, where some participants benefit more from VR and others benefit more from the real-world. Another study tested participants in a VR navigation task for mobile and tablet devices—Sea Hero Quest— and in similar navigation tasks in Paris (France) and London (UK) [20]. The virtual wayfinding task encompassed several skills involved in navigation, such as interpreting a map, planning a route with multiple stops, memorising the route, switching between an allocentric and an egocentric frame of reference, and monitoring and updating progress along the route. The real-world tasks took place in large-scale environments that covered a whole neighbourhood each – Covent Garden in London, and South of Montparnasse cemetery in Paris – and were designed to be close counterparts to the virtual task, with GPS tracking of participants. The study found that performance in the virtual task was predictive of real-world performance. Moreover, they found similar effects in both tasks (e.g. a male advantage). Establishing the ecological validity of the virtual test, provided critical support for the validity of inferences made from analysis of 4 million participants tested online via the Sea Hero Quest test [11,26–29] and the inferences from the tests about the features of environments that make them difficult to navigate [30].

While ecological validity might hold for younger participants it is unclear that testing older participants who are less familiar with VR would produce a similar pattern. A follow-up study to Coutrot et al (2019) had older participants completing the same virtual and real-world task (focusing on London, UK) [23], see Figure 1. For this older cohort, performance in Sea Hero Quest predicted real-world performance for levels with medium difficulty. Performance in the easy or difficult environments of the virtual task, however, did not predict real-world performance. While the result supports the usefulness of digital tests for older age groups, it also highlights that ecological validity is not an all or nothing quality, but needs to be explored for varying demographics, tasks and environmental conditions. A similar issue relates to test-retest reliability for tests used in the lab and real-world [31].

Another way to study navigation of real-world environments is to explore the planning process in real world situations [32–34]. This can be difficult due to the variability in people's knowledge of the environment. One way to address this is study experts who have obtained a sufficiently expert level of knowledge. One group of experts are London licensed taxi drivers [35]. By asking them to describe every street between one location and another as part of a plan it is possible to examine which steps through the street network take the most time to make and thus which are most demanding to compute. Examining these choices has revealed evidence for efficient hierarchical processing of the street network, with choices on boundaries to regions more quickly processes than would be expected [36] and evidence of pre-caching of the transition structure during an initial first consideration of the problem [34].

## 2. Virtual and simulated environments

Two recent studies have explored the extent to which the environment we live in and grow up in can impact spatial navigation performance. Barhorst-Cates and colleagues [37] explored the impact of growing up and living in Padua (Italy) versus Salt Lake City (USA). Salt lake city has a grid layout and very distinct distal cues from a mountain range. Padua has no strong distal cues, but rich proximal cues (landmarks on street corners) and non-grid, more randomly oriented street layout. The Padua participants were accurate at using proximal cues (e.g. landmarks in streets), which is the way most people navigate that city, and Salt Lake City individuals reported using more survey strategies (top down map-based). It is notable that some of the results ran counter to their hypotheses: Padua participants tended to take more shortcuts and pointed more accurately to familiar locations than Utah participants (Fig 1). A strength of the study was that tests were conducted both in relation to real-world choices and with standardised virtual tests of navigation.

The other study exploring this topic used Sea Hero Quest and examined the extent to whether growing up outside or inside cities impacts navigation ability [27]. Data from over 380,000 participants revealed that on average, growing up inside cities resulted in worse navigation skills across 38 countries. But it was countries with grid-like road layouts in their cities where this was most evident. Indeed, the more grid-like the city streets of a country the more negative the effect of growing up inside a city was (Figure 1). It seems plausible that growing up in cities with simplified layouts lowers the engagement brain structure during navigation having a knock-on effect on navigation ability. Recent evidence indicates that environments with more complex layouts (higher street intersection density) are associated with larger posterior hippocampal volume in older adults [38] and the posterior hippocampus is activated by complex junctions when navigating [39].

Another approach to exploring navigation behaviour and its neural basis has been the use of virtual and simulated environments to test healthy people during brain recordings or patients with brain lesions as they navigate environments. Combining brain imaging or neuropsychology with basic virtual environments dates back several decades [40–42].

Numerous studies have explored situations mimicking real-world scenarios in unfamiliar virtual environments [43–48]. The development of digital mapping tools and higher processing allowed for the creation of a virtual version of a real city (London UK) complete with simulated traffic and pedestrians, allowing examination of brain dynamics during navigation of a familiar environment [49,50]. However, creating such environments is expensive and difficult to update. Another effective way to test navigation for real-world environments has been the use of first-person view film simulations [39,51,52] or Google Street View (or similar images), in which participants can select route choices through the sequence of images matching the real-world [53–56]. These approaches have given insight into brain regions tracking the distance to the goal [51,55], representations of space [39,53,54], new learning [56] and memory for spaces travelled [52] (see Figure 2).

## 3. Mobile sensing

The widespread use of smartphones and wearable technology offers a new perspective on navigation in everyday settings. GPS tracking of thousands of vehicles has given insights into routing choices made, showing route choices can be surprisingly non-optimal [57–59]. One recent study used a tracking smartphone application to collect GPS data from 5,590 pedestrians in Boston and 8,790 in San Francisco, two cities with differing street network topology [60]. They analysed the deviations from shortest path between origin and destination in all the participants' trajectories, and they found that a vector-based navigation model was the best fit for the data. Longitudinal GPS tracking can be combined with spatial tasks. A study of children in a forager-horticulturalist population in Bolivia (the Tsimané) found no gender differences for either mobility or spatial ability [61]. They also found that in this population, schooling, which limits outdoor spatial exploration, was associated with poorer pointing accuracy. GPS trajectories are, of course, not the only form of mobile sensing. Data can also take other physical and digital forms, such as the structure of social networks derived from social media activity [62].

Smartphones also facilitate longitudinal experience sampling. One study geolocated participants based in New York and in Miami and prompted them to provide ecological momentary assessments of positive and negative affect through a questionnaire sent to their mobile phones [63]. They calculated daily variability (roaming entropy) in the participants' location, and they also used US Census Bureau data about the sociodemographic features of each location visited by participants (e.g. population density, median age, gender, etc) to calculate the entropy over a sociodemographic feature space. Both roaming and sociodemographic feature entropy were associated with day-to-day variation in positive affect. Combining mobile sensing approaches with resting-state fMRI scanning in the laboratory, they found that the effect was stronger for those participants with greater functional coupling of the hippocampus and striatum. Mobile sensing data has a lot of promise, but by itself it might not increase ecological validity [64]. Unforeseen events like traffic, social encounters, or accidents all can affect trajectory data, so carefully planned data analytics are of prime importance for this line of work [6].

## 4. Mobile brain recording

One element that facilitates the study of navigation as it unfolds in the real world is the increased portability of brain imaging technology, such as Functional Near-Infrared Spectroscopy (fNIRS) and mobile electroencephalography (EEG) [65–69]. One study used mobile EEG with a group of participants learning to navigate the Charlottenburg neighbourhood in Berlin (Germany) either with standard or with auditory landmark-based navigation instructions [70]. They extracted gait-related activity, saccade and blink events from sensor level data, and they found that brain potentials related to eye movements were indicative of higher cognitive processes being involved, which increased the processing of incoming information (during landmark-based instructions).

Mobile neuroimaging can also help increase the ecological validity of studies in the laboratory, because it allows participants navigating a virtual environment to move freely, rather than controlling their movement with controllers while stationary. Several studies have now employed mobile EEG in this way, and found that theta activity in the retrosplenial complex supports landmark-based correction during path integration [68] and anchors space to the body cardinal axes [71]. A promising variation of mobile EEG is mobile *intracranial* EEG, where recordings in the brain are made from patients awaiting surgical treatment for epilepsy (Fig 2). This has been used to discover boundary-anchored neural representations in the medial temporal lobe modulated by self -and other- location [72]. More recently [73] used iEEG to compare the brain dynamics of imagined and real-world navigation. This latter study found that intermittent hippocampal theta dynamics encode spatial information and segments navigational routes during real-world navigation and similar patterns in the absence of external stimuli during imagined navigation.

**Conclusion**
Navigation is a complex and dynamic phenomenon that is difficult to capture with traditional lab-based methods. This difficulty has led to a variety of ecological approaches. These span navigation tasks outside the lab, the validation of standardised virtual tests of navigation in real-world scenarios, adapting real-world navigation practices to experimental tasks and developing virtual environments that are increasingly close to real-world environments. Many experiments are now also paying special attention to the environments that different participants frequent or grew up in, and their effect on navigation behaviour. Finally, technological developments have driven this ecological effort forward, both through mobile sensing forms of data capture (e.g. GPS tracking) and through mobile neuroimaging (e.g. EEG). There is enormous potential in the combination of all these approaches. There are also many gaps, both methodologically and in terms of the varieties of real-world behaviour. There are only a few studies on collaborative wayfinding, or of navigation in crowded environments [74,75]. Existing research does not capture the wide range of environments that humans navigate (e.g. few to none experiments at sea), or the cultural diversity of wayfinding techniques [7]. In terms of technology, there is interesting potential in the use of immersive 360 video [76], and few fNIRS and mobile EEG studies have taken place in outdoor environments. We are hopeful that experimental work is increasingly ecological, and that it will gradually capture the beautiful complexity of navigation as it occurs in the real world.

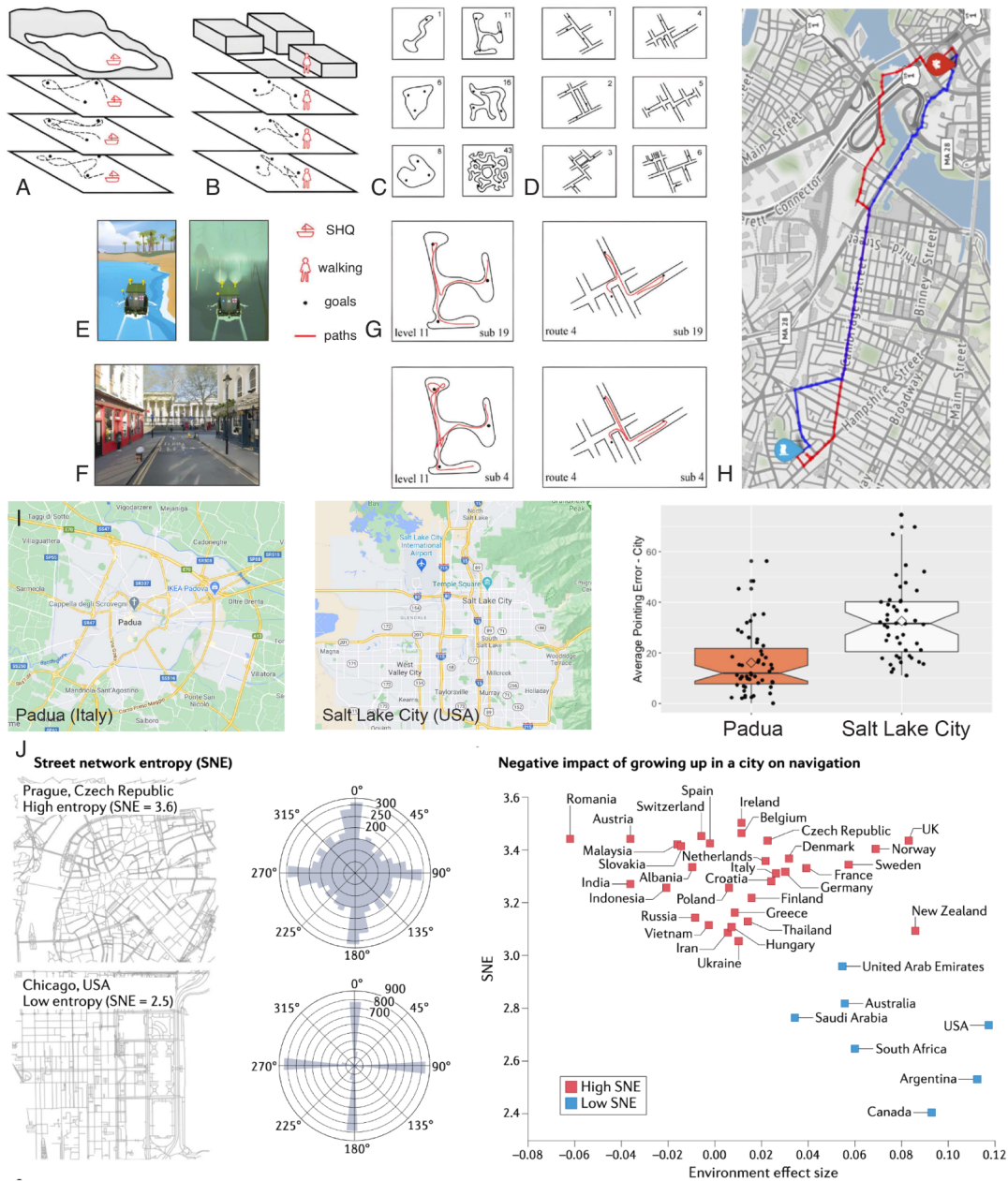

**Figure 1 - Virtual, Real, and Agentic Spatial Navigation.**
**A-** Example trajectories within the video game Sea Hero Quest. **B-** Example trajectories in a similar real-world wayfinding task. **C-** Six levels of Sea Hero Quest tested. **D-** Six real-world wayfinding routes tested. **E-** Two participant viewpoints from within Sea Hero Quest. **F-** Example participant viewpoint during route 1 of the real-world wayfinding task. **G-** Example trajectories taken by two different participants on level 11 of Sea Hero Quest and route 4 in the real-world task [23]. **H-** Asymmetry in human paths revealed in a massive dataset of GPS traces collected in Boston and San Francisco. Red paths, which start at red markers, differ from blue paths, which start at blue markers. The fact that chosen paths are statistically different when origin and destination are swapped is consistent with a vector-based navigation model for human path planning [60]. **I-** Participants living in Padua (Italy) had better spatial skills than those living in Salt Lake City (USA) (Barhorst-Cates, 2021). **J-** In countries where cities have less complex street networks (low Street Network Entropy), people raised in cities have worse spatial skills than those raised outside cities. In countries with cities more complex to navigate, this difference disappears [27]. The results presented in I and J support the idea that environmental demands shape spatial strategies and performances. Panels adapted with permission from [23,37,60,77].

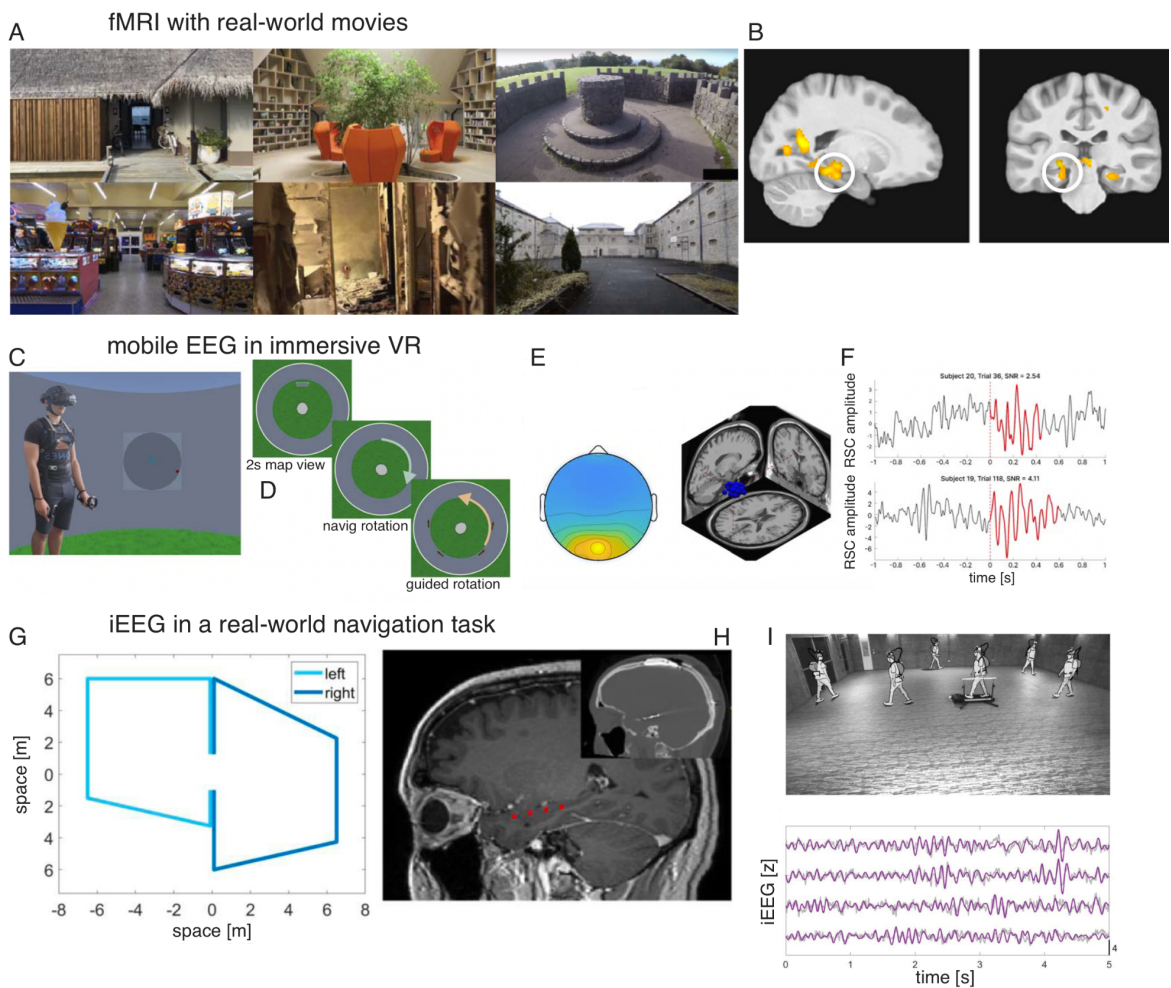

**Figure 2 - fMRI, mobile EEG, iEEG: different brain signals for different real-world aspects of spatial navigation**

**A-B:** fMRI and first-person movies to determine the contribution of different brain regions to spatial and aesthetic aspects of the built environment. A - Participants are presented with real-world movies in a MRI scanner, while asked to rate the valence or spatial complexity of each video. B- Hippocampal activity increased for videos that are later better remembered [52]. **C-F:** mobile EEG and immersive VR during goal-directed whole-body rotations show that theta bursts in the retrosplenial complex (RSC) encode both acceleration and alignment with the body's principal axes. C- Participants wore an immersive head-mounted display, along with a 64-channel EEG cap. D- Task protocol. E- EEG source analysis. Source-level dipoles were reconstructed from scalp EEG data and clustered across participants. F- Example of detected EEG bursts. Two selected theta bursts are shown, occurring near 90° angular transitions (0 on the x-axis indicates the moment participants passed through 90°) [71]. **G-**

**I:** Hippocampal theta oscillations during real-world and imagined navigation using motion capture and intracranial EEG recordings. G- Participants completed a spatial navigation task that involved the learning of two distinct routes. **H-** Electrodes in the hippocampus from an example participant are indicated in red. **I-** An illustrative time-lapse motion capture shows a selected participant navigating the rightward route. Exemplary broadband iEEG activity during real-world navigation [73]. Panels adapted with permission from [52,71,73].

**References**

1. Shamay-Tsoory SG, Mendelsohn A: **Real-Life Neuroscience: An Ecological Approach to Brain and Behavior Research**. *Perspect Psychol Sci* 2019, **14**:841–859.

2. Sonkusare S, Breakspear M, Guo C: **Naturalistic Stimuli in Neuroscience: Critically Acclaimed**. *Trends in Cognitive Sciences* 2019, **23**:699–714.

3. Nastase SA, Goldstein A, Hasson U: **Keep it real: rethinking the primacy of experimental control in cognitive neuroscience**. *NeuroImage* 2020, **222**:117254.

4. Ibanez A: **The mind's golden cage and cognition in the wild**. *Trends in Cognitive Sciences* 2022, **26**:1031–1034.

5. Maguire EA: **Does memory research have a realistic future?** *Trends in Cognitive Sciences* 2022, **26**:1043–1046.

6. Vigliocco G, Convertino L, Felice SD, Gregorians L, Kewenig V, Mueller MAE, Veselic S, Musolesi M, Hudson-Smith A, Tyler N, et al.: **Ecological Brain: Reframing the Study of Human Behaviour and Cognition**. 2024, doi:10.31234/osf.io/zr4nm.

7. Fernandez Velasco P, Spiers HJ: **Wayfinding across ocean and tundra: what traditional cultures teach us about navigation**. *Trends in Cognitive Sciences* 2024, **28**:56–71.

8. Wolbers T, Hegarty M: **What determines our navigational abilities?** *Trends in cognitive sciences* 2010, **14**:138–146.

9. Ekstrom AD, Spiers HJ, Bohbot VD, Rosenbaum RS: *Human spatial navigation*. Princeton University Press; 2018.

10. Patai EZ, Spiers HJ: **The versatile wayfinder: prefrontal contributions to spatial navigation**. *Trends in cognitive sciences* 2021, **25**:520–533.

11. Coutrot A, Silva R, Manley E, de Cothi W, Sami S, Bohbot VD, Wiener JM, Hölscher C, Dalton RC, Hornberger M, et al.: **Global Determinants of Navigation Ability**. *Current Biology* 2018, **28**:2861-2866.e4.

12. Spiers HJ, Coutrot A, Hornberger M: **Explaining World-Wide Variation in Navigation Ability from Millions of People: Citizen Science Project Sea Hero Quest**. *Topics in Cognitive Science* 2023, **15**:120–138.

13. Walkowiak S, Coutrot A, Hegarty M, Velasco PF, Wiener JM, Dalton RC, Hölscher C, Hornberger M, Spiers HJ, Manley E: **Cultural determinants of the gap between self-estimated navigation ability and wayfinding performance: evidence from 46 countries**. 2022, doi:10.1101/2022.10.19.512889.

14. Witmer BG, Bailey JH, Knerr BW, Parsons KC: **Virtual spaces and real world places: transfer of route knowledge**. *International Journal of Human-Computer Studies* 1996, **45**:413–428.

15. Waller D, Hunt E, Knapp D: **The Transfer of Spatial Knowledge in Virtual Environment Training**. *Presence: Teleoperators and Virtual Environments* 1998, **7**:129–143.

16. Richardson AE, Montello DR, Hegarty M: **Spatial knowledge acquisition from maps and from navigation in real and virtual environments**. *Memory & Cognition* 1999, **27**:741–750.

17. Montello DR, Waller D, Hegarty M, Richardson AE: **Spatial Memory of Real Environments, Virtual Environments, and Maps**. In *Human Spatial Memory*. . Psychology Press; 2004.

18. Sorita E, N'Kaoua B, Larrue F, Criquillon J, Simion A, Sauzéon H, Joseph P-A, Mazaux J-M: **Do patients with traumatic brain injury learn a route in the same way in real and virtual environments?** *Disability and Rehabilitation* 2013, **35**:1371–1379.

19. van der Ham IJM, Faber AME, Venselaar M, van Kreveld MJ, Löffler M: **Ecological validity of virtual environments to assess human navigation ability**. *Front Psychol* 2015, **6**.

20. Coutrot A, Schmidt S, Coutrot L, Pittman J, Hong L, Wiener JM, Hölscher C, Dalton RC, Hornberger M, Spiers HJ: **Virtual navigation tested on a mobile app is predictive of real-world wayfinding navigation performance**. *PLOS ONE* 2019, **14**:e0213272.

21. Hejtmanek L, Starrett M, Ferrer E, Ekstrom AD: **How Much of What We Learn in Virtual Reality Transfers to Real-World Navigation?** *Multisensory Research* 2020, **33**:479–503.

22. Bonavita A, Teghil A, Pesola MC, Guariglia C, D'Antonio F, Di Vita A, Boccia M: **Overcoming navigational challenges: A novel approach to the study and assessment of topographical orientation**. *Behav Res* 2022, **54**:752–762.

23. Goodroe S, Velasco PF, Gahnstrom CJ, Wiener J, Coutrot A, Hornberger M, Spiers HJ: **Predicting real-world navigation performance from a virtual navigation task in older adults**. *PLOS ONE* 2025, **20**:e0317026.

24. Dolce E, Gidari ML, Ruffo I, Longo A, Siciliano V, Canino S, Boccia M, D'Antonio F, Di Vita A, Raimo S, et al.: **How do individual differences in interoception influence navigation in virtual and real environments?** *Journal of Environmental Psychology* 2025, **108**:102814.

25. Roseman M, Elias U, Kletenik I, Ferguson MA, Fox MD, Horowitz Z, Marshall GA, Spiers HJ, Arzy S: **A neural circuit for spatial orientation derived from brain lesions**. *Cereb Cortex* 2024, **34**:bhad486.

26. Coutrot A, Lazar AS, Richards M, Manley E, Wiener JM, Dalton RC, Hornberger M, Spiers HJ: **Reported sleep duration reveals segmentation of the adult life-course into three phases**. *Nat Commun* 2022, **13**:7697.

27. Coutrot A, Manley E, Goodroe S, Gahnstrom C, Filomena G, Yesiltepe D, Dalton RC, Wiener JM, Hölscher C, Hornberger M, et al.: **Entropy of city street networks linked to future spatial navigation ability**. *Nature* 2022, **604**:104–110.

28. Coutrot A, Kievit RA, Ritchie SJ, Manley E, Wiener JM, Hölscher C, Dalton RC, Hornberger M, Spiers HJ: **Education Is Positively and Causally Linked With Spatial Navigation Ability Across the Lifespan**. *Open Mind* 2025, **9**:926–939.

29. Fernandez-Velasco P, Coutrot A, Oloye H, Wiener JM, Dalton RC, Holscher C, Manley E, Hornberger M, Spiers HJ: **No link between handedness and spatial navigation: evidence from over 400 000 participants in 41 countries**. *Proceedings of the Royal Society B: Biological Sciences* 2023, **290**:20231514.

30. Yesiltepe D, Fernández Velasco P, Coutrot A, Ozbil Torun A, Wiener JM, Holscher C, Hornberger M, Conroy Dalton R, Spiers HJ: **Entropy and a sub-group of geometric measures of paths predict the navigability of an environment**. *Cognition* 2023, **236**:105443.

31. Coughlan G, Puthusseryppady V, Lowry E, Gillings R, Spiers H, Minihane A-M, Hornberger M: **Test-retest reliability of spatial navigation in adults at-risk of Alzheimer's disease**. *PLOS ONE* 2020, **15**:e0239077.

32. T. Brunyé T, R. Mahoney C, L. Gardony A, A. Taylor H: **North is up(hill): Route planning heuristics in real-world environments**. *Memory & Cognition* 2010, **38**:700–712.

33. van Opheusden B, Kuperwajs I, Galbiati G, Bnaya Z, Li Y, Ma WJ: **Expertise increases planning depth in human gameplay**. *Nature* 2023, **618**:1000–1005.

34. Fernandez Velasco P, Griesbauer E-M, Brunec IK, Morley J, Manley E, McNamee DC, Spiers HJ: **Expert navigators deploy rational complexity–based decision precaching for large-scale real-world planning**. *Proceedings of the National Academy of Sciences* 2025, **122**:e2407814122.

35. Griesbauer E-M, Manley E, McNamee D, Morley J, Spiers H: **What determines a boundary for navigating a complex street network: evidence from London taxi drivers**. *The Journal of Navigation* 2022, **75**:15–34.

36. Griesbauer E-M, Fernandez Velasco P, Coutrot A, Wiener JM, Morley JG, McNamee D, Manley E, Spiers HJ: **London taxi drivers exploit neighbourhood boundaries for hierarchical route planning**. *Cognition* 2025, **256**:106014.

37. Barhorst-Cates EM, Meneghetti C, Zhao Y, Pazzaglia F, Creem-Regehr SH: **Effects of home environment structure on navigation preference and performance: A comparison in Veneto, Italy and Utah, USA**. *Journal of Environmental Psychology* 2021, **74**:101580.

38. Poudel G, Sachdev P, Soloveva M, Molina M, Schroers R-D, Li W, Knibbs LD, Catts V, Anstey K, Wen W, et al.: **Living in urban areas with highly interconnected streets is linked to greater posterior hippocampal volume in older adults**. 2025, doi:10.21203/rs.3.rs-7538181/v1.

39. Javadi A-H, Emo B, Howard LR, Zisch FE, Yu Y, Knight R, Pinelo Silva J, Spiers HJ: **Hippocampal and prefrontal processing of network topology to simulate the future**. *Nat Commun* 2017, **8**:14652.

40. Aguirre GK, Detre JA, Alsop DC, D'Esposito M: **The Parahippocampus Subserves Topographical Learning in Man**. *Cereb Cortex* 1996, **6**:823–829.

41. Maguire EA, Burgess N, Donnett JG, Frackowiak RS, Frith CD, O'Keefe J: **Knowing where and getting there: a human navigation network**. *Science* 1998, **280**:921–924.

42. Spiers HJ, Burgess N, Hartley T, Vargha-Khadem F, O'Keefe J: **Bilateral hippocampal pathology impairs topographical and episodic memory but not visual pattern matching**. *Hippocampus* 2001, **11**:715–725.

43. Brown TI, Ross RS, Keller JB, Hasselmo ME, Stern CE: **Which Way Was I Going? Contextual Retrieval Supports the Disambiguation of Well Learned Overlapping Navigational Routes**. *J Neurosci* 2010, **30**:7414–7422.

44. Slone E, Burles F, Iaria G: **Environmental layout complexity affects neural activity during navigation in humans**. *Eur J Neurosci* 2016, **43**:1146–1155.

45. Javadi A-H, Patai EZ, Marin-Garcia E, Margolis A, Tan H-RM, Kumaran D, Nardini M, Penny W, Duzel E, Dayan P: **Prefrontal dynamics associated with efficient detours and shortcuts: A combined functional magnetic resonance imaging and magnetoencenphalography study**. *Journal of cognitive neuroscience* 2019, **31**:1227–1247.

46. Brown TI, Gagnon SA, Wagner AD: **Stress Disrupts Human Hippocampal-Prefrontal Function during Prospective Spatial Navigation and Hinders Flexible Behavior**. *Curr Biol* 2020, **30**:1821-1833.e8.

47. Maxim P, Brown TI: **Differential representations of spatial environments in mPFC and hippocampus underpinning flexible navigation**. *NeuroImage* 2026, **327**:121736.

48. Lu Z, Julian JB, Aguirre GK, Epstein RA: **A Neural Compass in the Human Brain during Naturalistic Virtual Navigation**. *J Neurosci* 2025, **45**.

49. Spiers HJ, Maguire EA: **Spontaneous mentalizing during an interactive real world task: an fMRI study**. *Neuropsychologia* 2006, **44**:1674–1682.

50. Spiers HJ, Maguire EA: **The neuroscience of remote spatial memory: a tale of two cities**. *Neuroscience* 2007, **149**:7–27.

51. Howard LR, Javadi AH, Yu Y, Mill RD, Morrison LC, Knight R, Loftus MM, Staskute L, Spiers HJ: **The hippocampus and entorhinal cortex encode the path and Euclidean distances to goals during navigation**. *Current Biology* 2014, **24**:1331–1340.

52. Gregorians L, Patai Z, Velasco PF, Zisch FE, Spiers HJ: **Brain Dynamics during Architectural Experience: Prefrontal and Hippocampal Regions Track Aesthetics and Spatial Complexity**. *J Cogn Neurosci* 2026, **38**:381–405.

53. Chanales AJH, Oza A, Favila SE, Kuhl BA: **Overlap among Spatial Memories Triggers Repulsion of Hippocampal Representations**. *Curr Biol* 2017, **27**:2307-2317.e5.

54. Brunec IK, Bellana B, Ozubko JD, Man V, Robin J, Liu Z-X, Grady C, Rosenbaum RS, Winocur G, Barense MD, et al.: **Multiple Scales of Representation along the**

**Hippocampal Anteroposterior Axis in Humans**. *Current Biology* 2018, **28**:2129-2135.e6.

55. Patai EZ, Javadi A-H, Ozubko JD, O'Callaghan A, Ji S, Robin J, Grady C, Winocur G, Rosenbaum RS, Moscovitch M: **Hippocampal and retrosplenial goal distance coding after long-term consolidation of a real-world environment**. *Cerebral Cortex* 2019, **29**:2748–2758.

56. Ozubko JD, Campbell M, Verhayden A, Demetri B, Boyer MB, Sivashankar Y, Brunec I: **Hippocampal Signal Complexity Predicts Navigational Performance: Evidence From a Two-Week VR Training Program**. *Hippocampus* 2026, **36**:e70063.

57. Manley EJ, Addison JD, Cheng T: **Shortest path or anchor-based route choice: a large-scale empirical analysis of minicab routing in London**. *Journal of Transport Geography* 2015, **43**:123–139.

58. Lima A, Stanojevic R, Papagiannaki D, Rodriguez P, González MC: **Understanding individual routing behaviour**. *J R Soc Interface* 2016, **13**:20160021.

59. Teimouri F, Richter K-F, Hochmair HH: **Analysis of route choice based on path characteristics using Geolife GPS trajectories**. *Journal of Location Based Services* 2023, **17**:271–297.

60. Bongiorno C, Zhou Y, Kryven M, Theurel D, Rizzo A, Santi P, Tenenbaum J, Ratti C: **Vector-based pedestrian navigation in cities**. *Nature Computational Science* 2021, **1**:678–685.

61. Davis HE, Cashdan E: **Spatial cognition, navigation, and mobility among children in a forager-horticulturalist population, the Tsimané of Bolivia**. *Cognitive Development* 2019, **52**:100800.

62. Peer M, Hayman M, Tamir B, Arzy S: **Brain Coding of Social Network Structure**. *J Neurosci* 2021, **41**:4897–4909.

63. Heller AS, Shi TC, Ezie CEC, Reneau TR, Baez LM, Gibbons CJ, Hartley CA: **Association between real-world experiential diversity and positive affect relates to hippocampal–striatal functional connectivity**. *Nat Neurosci* 2020, **23**:800–804.

64. Holleman GA, Hooge ITC, Kemner C, Hessels RS: **The 'Real-World Approach' and Its Problems: A Critique of the Term Ecological Validity**. *Front Psychol* 2020, **11**.

65. Liu Y, Lu S, Liu J, Zhao M, Chao Y, Kang P: **A Characterization of Brain Area Activation in Orienteers with Different Map-Recognition Memory Ability Task Levels—Based on fNIRS Evidence**. *Brain Sciences* 2022, **12**.

66. Shi Y, Johnson C, Xia P, Kang J, Tyagi O, Mehta RK, Du J: **Neural Basis Analysis of Firefighters' Wayfinding Performance via Functional Near-Infrared Spectroscopy**. *Journal of Computing in Civil Engineering* 2022, **36**:04022016.

67. Ou S, Liu T, Liu Y: **Neural Mechanisms of the Impact of Rotated Terrain Symbols on Spatial Representation in Orienteers: Evidence from Eye-Tracking and Whole-Brain fNIRS Synchronization**. *Behavioral Sciences* 2025, **15**.

68. Naveilhan C, Zory R, Gramann K, Ramanoël S: **Theta Activity Supports Landmark-Based Correction of Naturalistic Human Path Integration**. *J Neurosci* 2025, **45**.

69. Djebbara Z, Huynh DC, Koselevs A, Chen Y, Fich LB, Gramann K: **Turning corners in built environments shifts spatial attention costs**. *NeuroImage* 2025, **322**:121549.

70. Wunderlich A, Gramann K: **Eye movement-related brain potentials during assisted navigation in real-world environments**. *European Journal of Neuroscience* 2021, **54**:8336–8354.

71. Naveilhan C, Ramanoël S: **Theta activity in the RSC anchors space to the body cardinal axes**. 2025, doi:10.64898/2025.12.02.691796.

72. Stangl M, Topalovic U, Inman CS, Hiller S, Villaroman D, Aghajan ZM, Christov-Moore L, Hasulak NR, Rao VR, Halpern CH, et al.: **Boundary-anchored neural mechanisms of location-encoding for self and others**. *Nature* 2021, **589**:420–425.

73. Seeber M, Stangl M, Vallejo Martelo M, Topalovic U, Hiller S, Halpern CH, Langevin J-P, Rao VR, Fried I, Eliashiv D, et al.: **Human neural dynamics of real-world and imagined navigation**. *Nat Hum Behav* 2025, **9**:781–793.

74. Bae C, Montello D, Hegarty M: **Wayfinding in pairs: comparing the planning and navigation performance of dyads and individuals in a real-world environment**. *Cogn Research* 2024, **9**:40.

75. Wang Z, Gabbana A, Toschi F: **Avalanches of choice: How stranger-to-stranger interactions shape crowd dynamics**. *Proceedings of the National Academy of Sciences* 2026, **123**:e2528167123.

76. Mazumder R, Spiers HJ, Ellard CG: **Exposure to high-rise buildings negatively influences affect: evidence from real world and 360-degree video**. *Cities & Health* 2022, **6**:1081–1093.

77. Farzanfar D, Spiers HJ, Moscovitch M, Rosenbaum RS: **From cognitive maps to spatial schemas**. *Nat Rev Neurosci* 2023, **24**:63–79.

 **Further information on references of particular interest**

22. ** This study explores real world behaviour and range of lab-based tests with the same participants

27 ** This study examines the impact of real-world environments layouts experienced in early life on later spatial cognition using a sample of over 380,000 participants

34 ** This study uses computational methods to explore real world route planning in expert navigators

37 ** This article shows how the environment one lives in can impact navigational abelites and strategies by testing people in two cultures/countries.

48 ** This study applies decoding to directional information in fMRI data in a much more naturalistic experience than all prior studies.

60 ** This study used the GPS data from a running app to explore route choices of people in real-world cities

63 ** This study used a novel combination of MRI and geotagging in an app to explore experience in the real-world in relation to brain activity.

72 ** This article shows how the human medial temporal lobe oscillations carry information about boundaries, both when walking and when watching someone else approach boundaries

73 **This article shows the similarity of the medial termpoal temporal lobe oscilations when particopant imagine walking on a route in a room, or actually walk in the room